\documentstyle[aps,prb,multicol]{revtex}
\input{psfig}
\begin{document}
\title{Non-equilibrium current noise in mesoscopic disordered 
SNS junctions}  
\author{Y. Naveh and D.V. Averin} 
\address{Department of Physics and Astronomy, SUNY at Stony Brook, 
Stony Brook NY 11794-3800}
\maketitle

\begin{abstract}
Current noise in superconductor-normal metal-superconductor (SNS)
junctions is calculated within the scattering theory of multiple
Andreev reflections (MAR). It is shown that the noise exhibits
subharmonic gap singularities at $eV=2\Delta/n$, $n=1,2,\, ... \, $
both in single-mode junctions with arbitrary transparency $D$ and in
multi-mode disordered junctions. The subharmonic structure is
superimposed with monotonic increase of the effective transferred
charge $q^*=S_I(0)/2I$ with decreasing bias voltage. Other features of
the noise include a step-like increase of $q^*$ in junctions with
small $D$, and a divergence $S_I(0) \propto V^{-1/2}$ at small
voltages and excess noise $S_{ex} = 2eI_{ex}$, where $I_{ex}$ is the
excess current, at large voltages, in junctions with diffusive
transport.
\end{abstract}
\begin{multicols}{2}
\vspace{-1cm}
The process of multiple Andreev reflections (MAR) in Josephson 
junctions with large electron transparency $D$ leads to a very 
non-trivial behavior of their transport properties, most 
notably, to the subharmonic gap singularities in the dc 
current $I$ at voltages $eV=2\Delta/n$, $n=1,2,\, ...\, $, where 
$\Delta$ is the superconducting gap -- see, e.g., [1] and 
references therein. In the past few years, experiments with atomic 
point contacts fabricated by the controllable break junction 
technique \cite{b2,b3,b4,Scheer} have demonstrated a remarkable 
level of agreement between theory and experiment in the description 
of the subgap structure in the dc current \cite{b4,Scheer}. 

The noise properties of high-transparency Josephson junctions 
are also non-trivial \cite{b5,b6,b7}. The main qualitative feature 
of the noise in the MAR regime is the increase of the effective 
transferred charge $S_I(0)/2I$, where $S_I(\omega)$ is the spectral 
density of current noise, with decreasing bias voltage. When the 
voltage decreases, the number $n$ of Andreev reflections in MAR 
cycles giving the dominant contribution to transport increases as 
$2\Delta/eV$. Since each pair of Andreev reflections from the two 
superconducting electrodes transfers the charge $2e$ across the 
junction, the effective noise charge increases as 
$2\Delta/V$.\cite{b6,b8,b9} 

Despite the fact that considerable experimental effort has been 
focused recently on measuring 
noise in diffusive superconductor-normal 
metal-superconductor (SNS) structures\cite{b9,Jehl,Kozhevnikov}, 
the noise has been studied theoretically only in single-mode 
ballistic junctions with $D=1$, where the enhancement of $q^*$ 
manifests itself only at temperatures $T\simeq \Delta$ and 
disappears at $T\rightarrow 0$. The aim of this work was to study 
the current noise associated with MAR in multi-mode junctions with 
arbitrary mode transparencies. We find that at finite reflection
coefficients the large low-voltage noise exists even at zero 
temperature, it exhibits pronounced subgap structure,  and is again
characterized by the effective charge  
on the order of $2\Delta/V$. 

In what follows, we use the standard scattering approach to the 
description of MAR in voltage-biased short disordered Josephson 
junctions -- see, e.g., [13]. The 
junction is modeled as a normal constriction between two 
superconductors with energy gap $\Delta$. The constriction region 
is assumed to support $N$ propagating electron modes and is 
characterized by the $2N$ by $2N$ scattering matrix $S$. Our 
objective is to find the spectral density $S_I(\omega)$ of 
current fluctuations 
\begin{equation}
S_I(\omega) =  \int d\tau e^{i\omega \tau }
\overline{ \langle I(t)I(t+\tau)+I(t+\tau)I(t)\rangle} \, , 
\label{1} \end{equation}  
where the bar indicates averaging over the time $t$, $I(t)= 
\hat{I}(t)- \langle \hat{I}(t) \rangle$ is the deviation of the 
current from its average value, and $\langle\, ... \, \rangle$  
denotes averaging over the equilibrium distribution of 
quasiparticles incident on the junction. The operator $\hat{I}$ 
of the current in the normal constriction is expressed in the 
standard way through the electron operators:   
\begin{equation}
\hat{I} = \frac{-ie\hbar}{2m} \sum_{\sigma} (\Psi^{\dagger}_{\sigma}
\frac{\partial \Psi_{\sigma}}{\partial x} - \frac{\partial 
\Psi^{\dagger}_{\sigma} }{\partial x}  \Psi_{\sigma})\, .
\label{2} \end{equation}
  
Since the superconductors induce electron-hole correlations in 
the normal region, the electron operator should be written as in 
the Bogolyubov's transformation: 
\begin{equation} 
\Psi_{\sigma}=\sum_{\nu} (u_{\nu} b_{\nu, \sigma}+ 
\mbox{sign}(\sigma) v^*_{\nu} b^{\dagger}_{\nu, -\sigma} ) \, .
\label{3} \end{equation} 
The operators $b_{\nu,\sigma}$ describe quasiparticles with spin 
$\sigma$ incident on the junction. The quasiparticle index $\nu$ 
consists of the mode number $i$, energy $\varepsilon$, and the 
index $s=L,R$ indicating whether the quasiparticle is incident 
from the left or right electrode: $\nu= \{i,\, \varepsilon, \, s\}$.
Substituting Eq.\ (\ref{3}) into (\ref{2}) and then into (\ref{1}) we 
can take two steps in transforming the expression (\ref{1}) for 
the noise $S_I(\omega)$. As a first step, using the properties of 
the scattering matrix $S$ (unitarity and time reversal symmetry) we 
prove that $S_I(\omega)$ can be presented as a sum of independent 
contributions of different modes. Then, we obtain expression for 
the contribution of a single mode in terms of the amplitudes of 
the forward- and backward-propagating components of electron and 
hole wavefunctions $u,\, v$ in Eq.\ (\ref{3}). The 
four relevant amplitudes are the amplitudes $A_n,\, B_n$ of Andreev 
and normal reflection with energy $\varepsilon +2neV$ for 
quasiparticle incident at energy $\varepsilon$, and the amplitudes 
$C_n,\, D_n$ of the normal and Andreev transmission with energy 
$\varepsilon +(2n+1)eV$. All these amplitudes are determined by the 
by-now standard set of recurrence relations that describe the 
process of MAR -- see, e.g., [13]. In terms of this 
amplitudes the current noise of a single mode is: 
\end{multicols}
\widetext
\begin{eqnarray}
& & S_I(\omega) =  \frac{e^2}{2\hbar} \sum_{\pm  \omega}  
\sum_{n,n'}\sum_{m,m'} \int d\epsilon d\epsilon' \left(1-
|a(\epsilon)|^2 \right)
\left(1- |a(\epsilon')|^2 \right) \times \left\{  \left[
f(\varepsilon) 
\left(1 - f(\varepsilon') \right) +
f(\varepsilon') \left(1-f(\varepsilon) \right) \right] \delta_{n-n',m-m'} 
\right. \nonumber \\
& & \hspace{1cm} \times \left\{ \delta \left[ \varepsilon -\varepsilon' 
+2(n-n')eV \pm \omega \right]  
\left[ \left(A_n A_{n'}^* +a_{2n}a^*_{2n'} \bar{A}_n \bar{A}_{n'}^* - 
\left(1+a_{2n}a^*_{2n'} \right)B_n B_{n'}^* \right)
\right. \right. \nonumber \\ 
& & \hspace{2cm} \times 
\left( A_m^* A_{m'} +a_{2m}^* a_{2m'} \bar{A}_m^* \bar{A}_{m'} - 
\left( 1+a_{2m}^*a_{2m'} \right) B_m^* B_{m'} \right)  \nonumber \\ 
& & \hspace{2cm} \left. + \left( 1+a_{2n+1}a^*_{2n'+1} \right)
\left(1+a_{2m+1}^*a_{2m'+1} \right)
\left( D_n D_{n'}^*-C_n C_{n'}^* \right) \left( D_m^* D_{m'}-C_m^*
C_{m'} \right) \right]  \nonumber \\
& &  \hspace{1.5cm} + 2\delta \left( \varepsilon -\varepsilon' 
+2(n-n')eV -eV \pm \omega \right) 
\left[A_n^* C_{n'} + a^*_{2n}a_{2n'+1} \bar{A}_n^* C_{n'} + 
\left( 1+a_{2n}^*a_{2n'+1} \right) B_{n}^*D_{n'} \right]  \nonumber \\ 
& & \left. \hspace{2cm} \times \left[ A_m C_{m'}^* +a_{2m} a_{2m'+1}^*
\bar{A}_m C_{m'}^* +  
\left( 1+a_{2m}a_{2m'+1}^* \right) B_m D_{m'}^* \right] \right\}
\nonumber \\ 
& & \hspace{1.cm} +  \left[ f(\varepsilon) f(\varepsilon') +
\left( 1-f(\varepsilon) \right) \left(1-f(\varepsilon') \right)
\right] \delta_{n+n',m+m'} \times \left\{ \delta \left[ 
\varepsilon +\varepsilon' +2(n+n')eV \pm \omega \right] \right. 
\nonumber \\
& & \hspace{1.5cm}\times \left[A_n^* B_{n'}^* + B_n^* A_{n'}^* 
-a_{2n}^*a_{2n'}^* 
\left( \bar{A}_n^* B_{n'}^*+ B_n^*\bar{A}_{n'}^* \right) \right]
\left[A_mB_{m'} + B_m A_{m'} -a_{2m}a_{2m'} \left(\bar{A}_m B_{m'}+ 
B_m\bar{A}_{m'} \right) \right] \nonumber \\ 
& & \hspace{1.cm} + \delta \left[\varepsilon +\varepsilon' +2(n+n'+1)eV 
\pm \omega \right]  
\left(1-a_{2n+1}a_{2n'+1} \right) \left(1-a_{2m+1}^*a_{2m'+1}^*
\right) \left(D_n C_{n'}+C_n D_{n'} \right) \left(D_m^* C_{m'}^*+C_m^*
D_{m'}^* \right)  \nonumber \\
& & \hspace{1.5cm} + 2\delta \left[\varepsilon +\varepsilon' +2(n+n')eV 
+eV \pm \omega \right] 
\left[ A_n^* D_{n'}^* - a^*_{2n}a^*_{2n'+1} \bar{A}_n^* D_{n'}^*  
- \left(1-a_{2n}^*a^*_{2n'+1} \right) B_{n}^*C_{n'}^* \right] \nonumber \\ 
& & \hspace{3cm} \left. \left. \left[ A_m D_{m'} -a_{2m} a_{2m'+1}
\bar{A}_m D_{m'} -  
\left( 1-a_{2m}a_{2m'+1} \right) B_m C_{m'} \right] \right\} \right\}
\, . \label{4}  
\end{eqnarray} 
\begin{multicols}{2}
Here $a_k\equiv a(\varepsilon+keV)$ with $a(\varepsilon)=
\mbox{sign} (\varepsilon ) (|\varepsilon |-(\varepsilon^2- 
\Delta^2)^{1/2} )/\Delta$ being the amplitude of Andreev 
reflection from a superconducting electrode, $\bar{A}_n\equiv 
A_n+ \delta_{n,0}/a_0$, and $f(\varepsilon)$ is the equilibrium 
Fermi distribution of quasiparticles. All 
quantities with primed indices in Eq.\ (\ref{4}) are functions 
of $\varepsilon'$. 
  
Despite its apparent complexity, the structure of Eq.\ (\ref{4}) 
is quite simple. The four groups of terms in this equation 
correspond to four different contributions to noise coming from 
the correlations between quasiparticles of the same 
type/different types (e.g., electron--electron or electron--hole) 
incident from the same electrode/opposite electrodes. The 
first group represents the contribution of similar quasiparticles 
coming from the same electrode, the second -- similar 
quasiparticles from opposite electrodes, etc. Results of the 
numerical calculation of noise for zero frequency and temperature 
from Eq.\ (4) are shown in Fig.\ 1.      

We see that the main qualitative feature of the noise for all
transparencies $D$ not too close to 0 and 1 is the subharmonic gap
structure at voltages $eV=2\Delta/n$, $n=1,2,\, ...\, $ that is
similar to, but more pronounced than, the subgap structure in the 
average current (shown in Fig.\ 1 by dashed lines for comparison). 
Both structures arise from the quasiparticle trajectories in energy 
space which hit the two edges of the superconducting energy gap,
$\varepsilon =\pm \Delta$, leading to resonant transmission of
quasiparticles  
through the contact. With 
the normalization 
used for $S_I(0)$ in Fig.\ 1 the noise and current curves should 
coincide when the noise is mainly due to uncorrelated transport of 
Cooper pairs. Figure 1 shows 
that this is indeed the case for relatively small transparencies in
the voltage range $\Delta <eV<  
2\Delta$. At very small transparencies
$D\rightarrow 0$ the noise in the subgap region $eV<2\Delta$
disappears together with the current and the curves approach the usual
tunnel-Hamiltonian result for quasiparticle noise in Josephson tunnel
junctions \cite{b11,b12}. This can also be checked analytically from
Eq.\ (\ref{4}) for arbitrary frequency $\omega$ and temperature $T$.

\vspace{-.2cm}
\begin{figure}[tbh] 
\vspace{0cm}
\centerline{\hspace{-.1cm} \psfig{figure=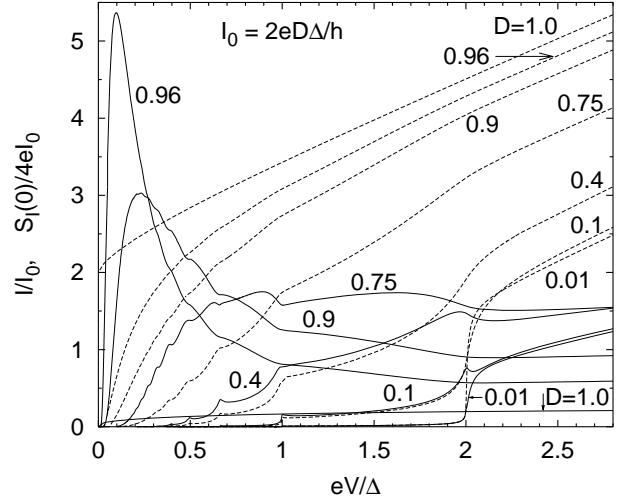,angle=-90,width=83mm}}
\narrowtext
\vspace{-.1 cm}
\caption{\hspace{0cm} Solid lines: Zero-temperature spectral density
of current fluctuations $S_I(0)$ as a function of the bias voltage $V$
in single-mode junctions with different electron transparencies
$D$. Dashed lines: the average dc current $I$. }
\label{1variousDs} 
\end{figure}

In the opposite limit of ballistic junction with $D=1$ Eq.\ 
(\ref{4}) reproduces the result for ballistic point contacts 
obtained previously \cite{b6}. At $\omega=0$, $T=0$ this result 
implies that, in agreement with Fig.\ 1, the noise is small 
for all bias voltages and approaches $16e^2\Delta/15\pi\hbar$ 
at $eV\gg \Delta$. \cite{b5} Figure 1 shows, however, that the 
curves with finite reflection coefficient $R=1-D\neq 0$ 
approach the ballistic limit $R=0$ in a non-trivial manner. 
As $R$ decreases, the noise develops a peak at small voltages. 
The peak becomes narrower and higher with decreasing $R$ and 
is shifted to smaller and smaller voltages, so that at 
$R\rightarrow 0$ it becomes infinitely narrow and disappears. 

This behavior of the noise at small bias voltages can be 
understood quantitatively using the concept of Landau-Zener 
transitions between the two subgap Andreev states developed 
before for the description of the average current \cite{b13}.  
The two states carry the currents $\pm (e\Delta /\hbar) \sin 
\varphi/2$, where $\varphi$ is the Josephson phase difference 
across the junction, $\dot{\varphi}=2eV/\hbar$. The origin 
of the noise in the regime of the Landau-Zener transitions is 
the stochastic quantum-mechanical nature of the transitions 
between the two states. In each period of the Josephson 
oscillations the junction either stays on one of the Andreev 
levels and carries the current of the same sign during the 
whole oscillation period, or makes the transition between the 
two states at $\varphi=\pi$ so that the current changes sign 
for the second half of the period. The first scenario is 
realized with the probability $p=\exp \{ -\pi R \Delta/eV \}$ 
and then the charge 
\[ Q_0 = \frac{e\Delta}{\hbar}  \int_0^{T_0} dt \sin \varphi/2 
=\frac{2\Delta}{V} \, , \]
where $T_0=\pi \hbar/eV$ is the period of the Josephson 
oscillations, is transferred through the junction. The second 
scenario has probability $1-p$ and no net charge is transferred 
in that case. Therefore statistics of charge $Q$ transferred 
through the junction during the large time interval $T\gg T_0$ 
is characterized by the binomial distribution with probability 
$p$. Definition (\ref{1}) then gives:
\begin{equation} 
S_I(0)= \frac{2(\langle Q^2 \rangle - \langle Q \rangle^2)}{ 
T} = \frac{8 e\Delta^2}{\pi \hbar V} p(1-p) \, . 
\label{5} \end{equation} 
The noise (\ref{5}) reaches maximum at small voltages $eV 
\simeq \pi R \Delta$ and its peak value increases with 
decreasing $R$ as $1/R$. One can check that Eq. (\ref{5}) 
describes accurately the low-voltage behavior of the curves 
with small $R$ in Fig.\ 1. 

Comparing Eq.\ (\ref{5}) to the expression for the average 
current in this regime\cite{b13}, $I=(2e\Delta /\pi \hbar)p$, 
we see that at small voltages, $eV\ll \pi R \Delta$, when 
$p\ll 1$ and both the average current and noise are suppressed, 
the  effective charge $q^*$ diverges as $1/V$, $q^*= S_I(0)/2I 
= 2\Delta/V$. The reason for this divergence is the coherent 
transfer of charge  within each cycle of MAR, the number $n$ of 
Andreev reflections in which is determined by the voltage, 
$n\simeq 2 \Delta/eV$. Each cycle transfers the charge $en$, 
and the current flows through the junction by rare random 
avalanches of Andreev reflections occurring with probability 
$p$ and transferring charge quanta $q^* = en = 2\Delta/V$ each. 
Therefore the noise in this regime is the shot noise of the 
large charge quanta $q^*$. 

From the approximate relation 
$q^*=en$, one can expect that $q^*$ could exhibit a step-like 
structure as a function of bias voltage due to discreteness of 
$n$. In junctions with large transparency $D \simeq 1$, 
however, different cycles of MAR contribute to transport 
simultaneously, and discreteness of $n$ is washed out. 
Nevertheless, when $D$ is small, the probability of the cycle 
of $n$ Andreev reflections is proportional to $D^n$, so the 
cycles with smallest possible $n$ dominate the transport. In 
this case the step structure in $q^*$ is pronounced, as can be 
seen from Fig.\ 2, where the average current and current noise 
are plotted as functions of voltage for two values of 
transparency $D\ll 1$ on a logarithmic scale. The noise curves 
in Fig.\ 2 follow closely the current with a shift that 
increases with decreasing voltage and reflects the increase of 
$q^*$. The inset in Fig.\ 2 shows directly the step structure in 
$q^*$ as a function of $2 \Delta /V$, and its gradual suppression 
with increasing $D$.   

\vspace{-0.2cm}
\begin{figure}[tbh] 
\vspace{0cm}
\centerline{\hspace{-.1cm} \psfig{figure=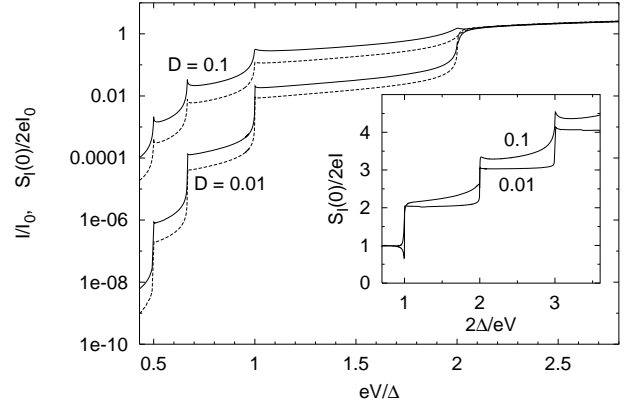,angle=-90,width=83mm}}
\narrowtext
\vspace{0.0cm}
\caption{ Current noise $S_I(0)$ and the average current $I$  
(same as in Fig.\ 1) on a logarithmic scale for two 
small values of the junction transparency $D$. Voltage 
dependence of the effective charge $q^*=S_I(0)/2I$ is shown 
in the inset.  \hspace{15cm}}
\label{2lowDs} 
\end{figure}

In the case of multimode junctions all the results obtained 
above should be averaged over the distribution of the mode 
transparencies. Below, we do this for the most important case 
of the SNS junction, where the N region is a disordered normal 
conductor with a large number of propagating modes and 
diffusive electron transport. Such a conductor is characterized 
by quasicontinuous Dorokhov's distribution of transparencies (see, 
e.g., [17] and references therein):  $\rho(D) = \pi \hbar 
G/2e^2D(1-D)^{1/2}$, where $G$ is the normal-state conductance 
of the N region. Averaging the numerical results for noise in 
single-mode junctions with this distribution we obtain the voltage
dependence of  
the non-equilibrium current noise in the SNS junctions, as shown in 
Fig.\ 3. We see that the averaging preserves the two qualitative 
features of the noise: subharmonic gap structure in the subgap 
region and excess noise at $eV\gg \Delta$. The new feature of 
$S_I(0)$ is the divergence at low voltages $eV\ll \Delta$. This 
divergence arises from the contribution of modes 
with small reflection coefficients and corresponds to the 
low-voltage noise peaks in the single-mode junctions. Averaging 
Eq.\ (\ref{5}) for these peaks with the Dorokhov's distribution     
we get the expression for the low-voltage noise in the SNS 
junction: 
\begin{equation} 
S_I(0) = {2 \Delta G} (\sqrt{2} - 1) 
\left( \frac{2 \Delta}{eV} \right)^{1/2} \, . 
\label{6} \end{equation}

Equation (\ref{6}) agrees with Fig.\ 3 and shows that at $eV\ll 
\Delta$ the noise diverges as $1/\sqrt{V}$. In presence of a 
finite energy relaxation (not accounted for in our scattering 
approach) this divergence would saturate at $eV\simeq \hbar 
\gamma$, where $\gamma$ is the relaxation rate.\cite{b15} 
Combining Eq.\ (\ref{6}) with the similar expression for the 
dc current\cite{b10}, $I=G(V\Delta/e)^{1/2}$, we find that in 
the diffusive junction the effective charge $q^*$ has the same 
$1/V$ voltage dependence as in the nearly ballistic single-mode 
junction (but with a different coefficient):
\begin{equation} 
q^*= (1-\frac{1}{\sqrt{2}}) \frac{2\Delta}{V} \, . 
\label{7} \end{equation}
Equation (\ref{7}) represents 
the main trend in the behavior of $q^*$ as a function of the 
bias voltage. The inset in Fig.\ 3 shows that on top of this 
linear growth of $q^*$ with $1/V$, it exhibits weak periodic 
oscillations associated with the subharmonic gap structure.  

\vspace{-0.2cm}
\begin{figure}[tbh] 
\vspace{0cm}
\centerline{\hspace{1cm} \psfig{figure=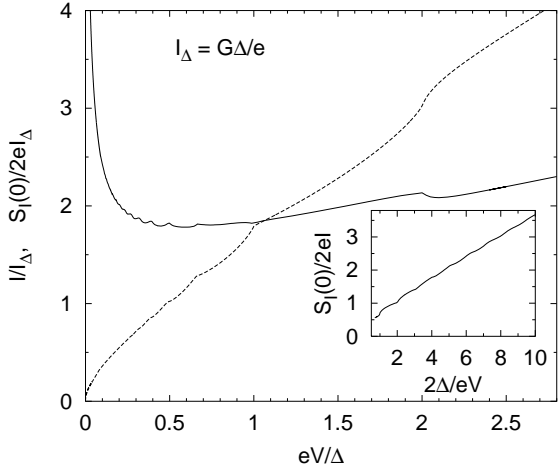,angle=-90,width=95mm}}
\narrowtext
\vspace{0.0cm}
\caption{ Current noise $S_I(0)$ (solid line) and the average current
$I$ (dashed line)  
of a short diffusive SNS junction as functions of the bias 
voltage $V$ at zero temperature. The inset shows the 
effective noise charge $q^*$ versus the inverse voltage. 
 \hspace{15cm}}
\label{3diffusive} 
\end{figure}

Next we discuss the excess noise. It can be seen from Eq.\ (4) 
and easily understood 
on physical grounds, that the leading contribution to the noise 
at large voltages $eV\gg \Delta$ coincides with the noise of the 
normal junctions\cite{de Jong}, $S_I(0)=(2e^3/\pi\hbar)VD(1-D)$ in the 
single-mode and $S_I(0)=2eGV/3$ in the multimode diffusive 
junction. In addition to this leading normal term that increases 
with voltage, there is an extra voltage-independent contribution 
$S_{ex}$ to the large-voltage noise.  This ``excess'' noise can be 
seen in Figures 1 and 3 and is directly related to the process of 
Andreev reflection. Indeed, one contribution to $S_{ex}$ comes from 
the quasiparticles incident on the superconducting electrodes at 
energies $\varepsilon$ outside the energy gap, $\mid \varepsilon 
\mid > \Delta$. For such quasiparticles, the probability of Andreev 
reflection is less than one and the random process of Andreev 
reflection/transmission introduces extra noise in addition to the 
noise associated with random normal reflection/transmission due 
to scattering in the normal region. Another contribution to 
$S_{ex}$ is due to quasiparticles inside the energy gap, 
$\mid \varepsilon \mid < \Delta$. Since each quasiparticle that is 
Andreev reflected in this energy range transfers the charge $2e$ 
(instead of the ``normal'' $e$), Andreev reflection amplifies the 
randomness of the normal scattering and increases the noise beyond 
its normal-state value. 

An expression for the excess noise in a single-mode 
junction as a function of transparency $D$ can be obtained from 
Eq.\ (\ref{4}) using the fact that only MAR processes with one 
Andreev reflection contribute to the noise in the $eV\gg \Delta$ 
limit. Averaging the single-mode expression over the Dorokhov's 
distribution we find the excess noise of the diffusive SNS 
junctions: 
\begin{equation} 
S_{ex} =2G\Delta (\frac{\pi^2}{4}-1) =2eI_{ex} \, . 
\label{8} \end{equation} 
The last equality in Eq.\ (\ref{8}) follows from the known 
expression for the excess current of such junctions \cite{b16}. 
We see that due to a curious numerical coincidence, the excess 
noise and current are related by a regular Schottky-like relation, 
in contrast to the leading normal contribution to the noise which 
is a factor of $1/3$ smaller than the Schottky expression. 

In summary, we have calculated the non-equilibrium current noise 
in the MAR regime in voltage biased short multimode Josephson 
junctions with arbitrary electron transmission properties. 
The noise exhibits several qualitative features associated with 
the coherent charge transfer by MAR cycles: subharmonic gap 
structure, ``quantization'' of the effective charge $q^*$, 
pronounced zero-temperature singularity at low bias voltages, and 
excess noise at large voltages.

\vspace*{1ex}

We would like to thank M.Devoret, D. Esteve, K.K. Likharev, 
V. Shumeiko, and C. Urbina for useful discussions of the 
results. We also note that when the manuscript was being prepared we
learned about a recent preprint \cite{b17} where the noise  
in the MAR regime was calculated for a single-mode junction.    
This work was supported in part by ONR grant \# N00014-95-1-0762.

\end{multicols}
\end{document}